\PassOptionsToPackage{unicode}{hyperref}
\PassOptionsToPackage{hyphens}{url}
\PassOptionsToPackage{dvipsnames,svgnames,x11names}{xcolor}
\documentclass[
]{article}
\usepackage{amsmath,amssymb}
\usepackage{lmodern}
\usepackage{iftex}
\ifPDFTeX
  \usepackage[T1]{fontenc}
  \usepackage[utf8]{inputenc}
  \usepackage{textcomp} 
\else 
  \usepackage{unicode-math}
  \defaultfontfeatures{Scale=MatchLowercase}
  \defaultfontfeatures[\rmfamily]{Ligatures=TeX,Scale=1}
\fi
\IfFileExists{upquote.sty}{\usepackage{upquote}}{}
\IfFileExists{microtype.sty}{
  \usepackage[]{microtype}
  \UseMicrotypeSet[protrusion]{basicmath} 
}{}
\makeatletter
\@ifundefined{KOMAClassName}{
  \IfFileExists{parskip.sty}{%
    \usepackage{parskip}
  }{
    \setlength{\parindent}{0pt}
    \setlength{\parskip}{6pt plus 2pt minus 1pt}}
}{
  \KOMAoptions{parskip=half}}
\makeatother
\usepackage{xcolor}
\usepackage{graphicx}
\makeatletter
\def\maxwidth{\ifdim\Gin@nat@width>\linewidth\linewidth\else\Gin@nat@width\fi}
\def\maxheight{\ifdim\Gin@nat@height>\textheight\textheight\else\Gin@nat@height\fi}
\makeatother
\setkeys{Gin}{width=\maxwidth,height=\maxheight,keepaspectratio}
\makeatletter
\def\fps@figure{htbp}
\makeatother
\providecommand{\tightlist}{%
  \setlength{\itemsep}{0pt}\setlength{\parskip}{0pt}}
\NewDocumentCommand\citeproctext{}{}
\NewDocumentCommand\citeproc{mm}{%
  \begingroup\def\citeproctext{#2}\cite{#1}\endgroup}
\makeatletter
 \let\@cite@ofmt\@firstofone
 \def\@biblabel#1{}
 \def\@cite#1#2{{#1\if@tempswa , #2\fi}}
\makeatother
\newlength{\cslhangindent}
\newlength{\csllabelwidth}
\newenvironment{CSLReferences}[2] 
 {\begin{list}{}{%
  \setlength{\itemindent}{0pt}
  \setlength{\leftmargin}{0pt}
  \setlength{\parsep}{0pt}
  \ifodd #1
   \setlength{\leftmargin}{\cslhangindent}
   \setlength{\itemindent}{-1\cslhangindent}
  \fi
  \setlength{\itemsep}{#2\baselineskip}}}
 {\end{list}}
\usepackage{calc}

\ifLuaTeX
\usepackage[bidi=basic]{babel}
\else
\usepackage[bidi=default]{babel}
\fi
\babelprovide[main,import]{american}

\def\languageshorthands#1{}
\ifLuaTeX
  \usepackage{selnolig}  
\fi
\IfFileExists{bookmark.sty}{\usepackage{bookmark}}{\usepackage{hyperref}}
\IfFileExists{xurl.sty}{\usepackage{xurl}}{} 
\hypersetup{
  pdftitle={21cmSense v2: A modular, open-source 21 cm sensitivity
calculator},
  pdfauthor={Steven G. Murray, Jonathan Pober, Matthew Kolopanis},
  pdflang={en-US},
  colorlinks=true,
  linkcolor={Maroon},
  filecolor={Maroon},
  citecolor={Blue},
  urlcolor={Blue},
  pdfcreator={LaTeX via pandoc}}

\title{21cmSense v2: A modular, open-source 21 cm sensitivity
calculator}

\usepackage[affil-it]{authblk}
\usepackage{orcidlink}
\author[1%
  *%
  \ensuremath\mathparagraph]{Steven G. Murray%
    \,\orcidlink{0000-0003-3059-3823}\,%
    }
\author[2%
  *%
  ]{Jonathan Pober%
    \,\orcidlink{0000-0002-3492-0433}\,%
    }
\author[3%
  ]{Matthew Kolopanis%
    \,\orcidlink{0000-0002-2950-2974}\,%
    }

\affil[1]{Scuola Normale Superiore, Italy}
\affil[2]{Department of Physics, Brown University, Providence, RI, USA}
\affil[3]{School of Earth and Space Exploration, Arizona State
University, Tempe, AZ, USA}
\affil[$\mathparagraph$]{Corresponding author: %
}
\affil[*]{These authors contributed equally.}
\date{18 January 2024}

\begin{document}
\maketitle

\section{Summary}\label{summary}

The 21cm line of neutral hydrogen is a powerful probe of the
high-redshift universe (Cosmic Dawn and the Epoch of Reionization), with
an unprecedented potential to inform us about key processes of early
galaxy formation, the first stars and even cosmology and structure
formation (\citeproc{ref-Liu2020}{Liu \& Shaw, 2020}), via intensity
mapping. It is the subject of a number of current and upcoming
low-frequency radio experiments, including the MWA
(\citeproc{ref-mwa}{Tingay et al., 2013}), LOFAR
(\citeproc{ref-lofar}{van Haarlem et al., 2013}), HERA
(\citeproc{ref-hera}{DeBoer et al., 2017}) and the SKA
(\citeproc{ref-Pritchard2015}{Pritchard et al., 2015}), which complement
the detailed information concerning the brightest sources in these early
epochs from powerful optical and near-infrared telescopes such as the
JWST (\citeproc{ref-jwst}{Castellano et al., 2022}).

21cmSense is a Python package that provides a modular framework for
calculating the sensitivity of these experiments, in order to enhance
the process of their design. This paper presents version v2.0.0 of
21cmSense, which has been re-written from the ground up to be more
modular and extensible, and to provide a more user-friendly interface --
as well as converting the well-used legacy package, presented in
(\citeproc{ref-Pober2013}{Pober et al., 2013},
\citeproc{ref-Pober2014}{2014}) from Python 2 to 3.

21cmSense can compute sensitivity estimates for both map-making
(\citeproc{ref-fhd}{Barry et al., 2019}) and delay-spectrum
(\citeproc{ref-Parsons2012}{Parsons et al., 2012}) approaches to
power-spectrum estimation. The full sensitivity calculation is rather
involved and computationally expensive in its most general form, however
21cmSense uses a few key assumptions to accelerate the calculation:

\begin{enumerate}
\def\labelenumi{\arabic{enumi}.}
\tightlist
\item
  Each baseline (pair of antennas) in the interferometer intrinsically
  measures a dense blob of 2D spatial Fourier modes of the sky intensity
  distribution, centred at a particular Fourier coordinate \((u,v)\)
  given by the displacement vector between the antennas forming the
  baseline, and covering an area in this \((u,v)\)-space that is given
  by the Fourier-transform of the primary beam of the instrument. The
  Fourier-space representation of the sky is thus built up by collecting
  many baselines that cover the so-called ``\((u,v)\)-plane''.
  \texttt{21cmSense} approximates this process of synthesising many
  baselines by nearest-grid-point interpolation onto a regular grid in
  the \((u,v)\)-plane. Furthermore, importantly the \((u,v)\)-grid is
  chosen to have cells that are comparable to the instrument's
  Fourier-space beam size, so that a particular baseline essentially
  measures a single cell in the grid, and no more. This maximizes
  resolution while keeping the covariance between cells small. This
  removes the need for tracking the full covariance between cells, and
  also removes the need to perform a beam convolution, which can be
  expensive.
\item
  We do not consider flagging of visibilities due to RFI and other
  systematics, which can complicate the propagation of uncertainties.
\end{enumerate}

Some of the key new features introduced in this version of 21cmSense
include:

\begin{enumerate}
\def\labelenumi{\arabic{enumi}.}
\tightlist
\item
  Simplified, modular library API: the calculation has been split into
  modules that can be used independently (for example, a class defining
  the \texttt{Observatory}, the \texttt{Observation} and the
  \texttt{Sensitivity}). These can be used interactively via Jupyter
  (\citeproc{ref-jupyter}{Kluyver et al., 2016}) or other interactive
  interfaces for Python, or called as library functions in other code.
\item
  Command-line interface: the library can be called from the
  command-line, allowing for easy scripting and automation of
  sensitivity calculations.
\item
  More accurate cosmological calculations using \texttt{astropy}
  (\citeproc{ref-astropy}{Astropy Collaboration et al., 2018};
  \citeproc{ref-Robitaille2013}{Robitaille et al., 2013})
\item
  Improved documentation and examples, including a Jupyter notebook that
  walks through the calculation step-by-step.
\item
  Generalization of the sensitivity calculation. The
  \texttt{Sensitivity} class is an abstract class from which the
  sensitivity of differing summary statistics can be defined. Currently,
  its only implementation is the \texttt{PowerSpectrum} class, which
  computes the classic sensitivity of the power spectrum. However, the
  framework can be extended to other summaries, for example wavelets
  (\citeproc{ref-Trott2016a}{Trott, 2016}).
\item
  Improved speed: the new version of 21cmSense is significantly faster
  than the legacy version, due to a number of vectorization improvements
  in the code.
\item
  Built-in profiles for several major experiments: MWA, HERA and SKA-1.
  These can be used as-is, or as a starting point for defining a custom
  instrument.
\end{enumerate}

\begin{figure}
\centering
\includegraphics{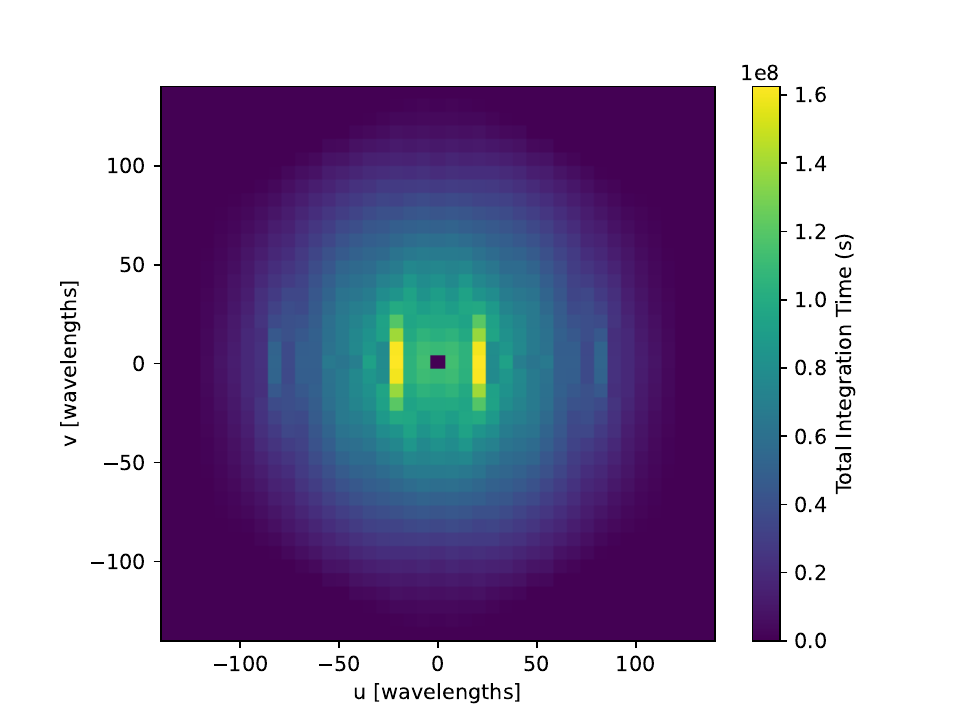}
\caption{Sampling of the \((u,v)\)-plane for the HERA experiment during
a full year of observations.\label{uvsampling}}
\end{figure}

\begin{figure}
\centering
\includegraphics{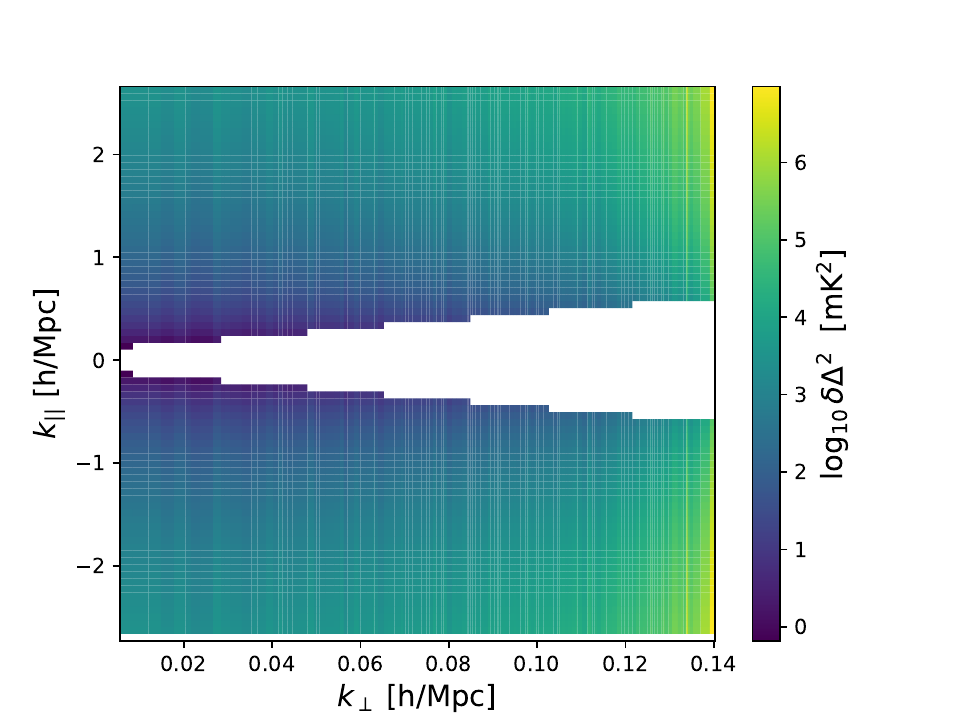}
\caption{Predicted noise-power of 1000 hours (one year) of HERA
observations, as a function of perpendicular and line-of-sight fourier
scale. The noise-power is represented for each
\(k\)-model.\label{sense}}
\end{figure}

An example of the predicted sensitivity of the HERA experiment after a
year's observation at \(z=8.5\) is shown in Figure \ref{sense},
corresponding to the sampling of the \((u,v)\)-grid shown in Figure
\ref{uvsampling}. The sensivity here is represented as a ``noise power''
(i.e.~the contribution to the power spectrum from thermal noise). This
figure also demonstrates that the new 21cmSense is capable of producing
sensitivity predictions in the cylindrically-averaged 2D power spectrum
space, which is helpful for upcoming experiments.

\section{Statement of need}\label{statement-of-need}

\texttt{21cmSense} provides a simple interface for computing the
expected sensitivity of radio interferometers that aim to measure the
21cm line of neutral hydrogen. This field is growing rapidly, with a
number of experiments currently underway or in the planning stages.
Historically, \texttt{21cmSense} has been a trusted tool for the design
of these experiments (\citeproc{ref-Greig2020}{Greig et al., 2020};
\citeproc{ref-Pober2013}{Pober et al., 2013},
\citeproc{ref-Pober2014}{2014}) and for forecasting parameter
constraints (\citeproc{ref-Greig2015}{Greig \& Mesinger, 2015},
\citeproc{ref-Greig2017}{2017}, \citeproc{ref-Greig2018}{2018}). This
overhauled, modularized version of \texttt{21cmSense} provides a more
user-friendly interface, improved performance, and the extensibility
required for the next generation, as evidenced by its usage in the
literature (\citeproc{ref-Breitman2024}{Breitman et al., 2024};
\citeproc{ref-Schosser2024}{Schosser et al., 2024}).

\section{Acknowledgements}\label{acknowledgements}

We acknowledge helpful conversations with Danny Jacobs.

\section*{References}\label{references}
\addcontentsline{toc}{section}{References}

\phantomsection\label{refs}

\end{document}